\documentclass[12pt]{article}
\usepackage[centertags]{amsmath}

\usepackage{amssymb}
\usepackage{latexsym}

\allowdisplaybreaks

\newcommand{\al}{\ensuremath{\alpha}}
\newcommand{\be}{\ensuremath{\beta}}
\newcommand{\La}{\ensuremath{\Lambda}}
\newcommand{\la}{\ensuremath{\lambda}}

\newcommand{\Ga}{\ensuremath{\Gamma}}
\newcommand{\de}{\ensuremath{\delta}}
\newcommand{\De}{\ensuremath{\Delta}}
\newcommand{\Up}{\ensuremath{\Upsilon}}

\newcommand{\ka}{\ensuremath{\kappa}}
\newcommand{\si}{\ensuremath{\sigma}}
\newcommand{\Si}{\ensuremath{\Sigma}}
\newcommand{\te}{\ensuremath{\theta}}
\newcommand{\ep}{\ensuremath{\varepsilon}}
\newcommand{\sfrac}[2]{\ensuremath{{\scriptstyle \frac{#1}{#2}}}}

\begin{document}

\setlength{\abovedisplayskip}{10pt plus 3pt minus 9pt}
\setlength{\belowdisplayskip}{10pt plus 3pt minus 9pt}
\setlength{\abovedisplayshortskip}{0pt plus 3pt}
\setlength{\belowdisplayshortskip}{5pt plus 3pt minus 4pt}

\title{\bf Supergravity and IOSp$\mathbf{(3,1|4)}$ gauge theory}
\author{T.G.\ Philbin\thanks{E-mail: tgp@ucc.ie} \\
 \small \it  Department of Mathematics, University College Cork, Cork, Ireland}

\date{}

\maketitle

\begin{abstract}
A new formulation of simple $D=4$ supergravity in terms of the geometry of superspace is presented. The formulation is derived from the gauge theory of the inhomogeneous orthosymplectic group IOSp$(3,1|4)$ on a $(4,4)$-dimensional base supermanifold by imposing constraints and taking a limit. Both the constraints and the limiting procedure have a clear {\it a priori} physical motivation, arising from the relationship between IOSp$(3,1|4)$ and the super Poincar\'{e} group. The construction has similarities with the space-time formulation of Newtonian gravity.
\end{abstract}

\noindent
{\small  PACS:  04.65.+e; 11.15.-q }

\section{Introduction}
The fact that general relativity is a theory of the dynamics of space-time geometry leads to the supposition that its supersymmetric version may also have a simple geometrical formulation, where the geometry involved is that of superspace. This does not appear to be the case, however. The analogue in superspace of general relativity, known as gauge supersymmetry~\cite{nat1}, is certainly not supergravity, so one must consider a more complicated superspace geometry. Nath and Arnowitt~\cite{nat2} obtained supergravity from gauge supersymmetry by contracting the tangent-space group OSp$(3,1|4)$ of the (4,4)-dimensional supermanifold to its SO(3,1) subgroup and taking a limit, while the standard Wess--Zumino formulation~\cite{wes1,wes2} requires the imposition of constraints on the superspace torsion. A drawback of both these formalisms is that they have elements, viz the limiting procedure of the former and the constraints of the latter, that have no clear {\it a priori} physical motivation.  In this article we seek a superspace formulation of supergravity that, in contrast to the Nath--Arnowitt and Wess--Zumino formulations, derives from an {\it a priori}, physically motivated principle. 

We shall obtain simple $D=4$ supergravity from the gauge theory of the inhomogeneous orthosymplectic group IOSp$(3,1|4)$ on a $(4,4)$-dimensional base supermanifold. This is achieved by imposing constraints and taking a limit, however both the constraints and the limiting procedure are determined at the outset by physical considerations. We point out that the resulting superspace geometry has similarities with Cartan's space-time formulation of Newtonian gravity~\cite{mis}. Indeed, the method used here to derive supergravity from IOSp$(3,1|4)$ gauge theory can also be used to derive Cartan's picture of Newtonian gravity from general relativity, when the latter is formulated as Poincar\'{e} gauge theory.

As explained above, the work described here was undertaken because of drawbacks of the Nath--Arnowitt and Wess--Zumino approaches to supergravity. We should mention a constraint-free, superfield formulation of simple $D=4$ supergravity due to Siegel and Gates~\cite{sie}, and Ogievetsky and Sokatchev (see~\cite{nie81} and references therein). This superfield approach has the advantage of being analogous to the superfield formulation of supersymmetric Yang--Mills theory.

Section~2 uses superspace to explore the relationship between IOSp$(3,1|4)$ and the super Poincar\'{e} group; we find that the latter can be
obtained from the former by the imposition of constraints and the taking of a limit. In Section~3 we derive the super Lie algebra of IOSp$(3,1|4)$ and show that one obtains the super Poincar\'{e} algebra by applying the constraints and limit of Section~2. Section~4 provides the necessary background in IOSp$(3,1|4)$ gauge theory. In Section~5 we apply the constraints and limit of Section~2 to the gauge potential and curvature of IOSp$(3,1|4)$ gauge theory and thereby derive supergravity. We then point out that our results make clear a geometrical relationship between supergravity and gauge supersymmetry that has similarities with the geometrical relationship between Newtonian gravity and general relativity.

\section{Superspace and the super Poincar\'{e} group}
Superspace is remarkable as a supermanifold in that its coordinates\footnote{Superspace coordinate indices are denoted by $\La=(\mu,\al,\dot{\al})$, orthosymplectic-frame indices by $A=(m,a,\dot{a})$; rules for super index positioning and manipulation are those of DeWitt~\cite{dew}. The space-time metric has signature $+2$ and the Infeld--van der Waerden symbols are $\si^{m}_{\ \ a\dot{a}}=\sfrac{i}{\sqrt{2}}(I,\vec{\si})$, where $\vec{\si}$ are the Pauli matrices.} $Z^\La=(x^\mu,\te^\al,\te^{\dot{\al}})$ have different mass dimensions $[Z^\La]$:
\begin{equation} \label{Coord}
[x^\mu]=-1, \qquad [\te^\al]=[\te^{\dot{\al}}]=-\sfrac{1}{2}.
\end{equation}
The reason for this is that the scale of the supersymmetry generators $Q_\al,Q_{\dot{\al}}$ is chosen to avoid introducing a physically irrelevant constant into the supersymmetry algebra~\cite{wes3}. Eqn.~(\ref{Coord}) has the consequence (though this is sometimes ignored) that the canonical metric in flat Riemannian superspace~\cite{dew} cannot have dimensionless components; for distance to be real with units of $x^\mu$ the canonical metric is \vspace{2mm}
\begin{equation} \label{Met}
_\La H_\Pi=\left(\begin{array}{ccc} \eta_{\mu\nu} & 0 & 0 \\
0 & k^2\ep_{\al\be} & 0 \\
0 & 0 & -k^2\ep_{\dot{\al}\dot{\be}} \end{array}\right) \quad
\Rightarrow \quad ^\La H^\Pi=\left(\begin{array}{ccc} \eta^{\mu\nu} & 0 & 0 \\
0 & -\frac{1}{k^2}\ep^{\al\be} & 0 \\
0 & 0 & \frac{1}{k^2}\ep^{\dot{\al}\dot{\be}} \end{array}\right), \vspace{2mm}
\end{equation}
where $k$ is a real constant with $[k]=-\sfrac{1}{2}$. We then have
\[
H(\De Z,\De Z)=\De Z^\La\,_\La H_\Pi\,\De Z^\Pi=\De x_\mu\,\De x^\mu+k^2\De\te_\al\,\De\te^\al-k^2\De\te_{\dot{\al}}\,\De\te^{\dot{\al}},
\]
so distance squared is real with units $[x^\mu]^2=-2$.

The group of coordinate transformations that leaves (\ref{Met}) unchanged is the analogue in superspace of the Poincar\'{e} group in space-time; it is the inhomogeneous orthosymplectic group IOSp$(3,1|4)$, i.e.\ the orthosympectic group OSp$(3,1|4)$ plus translations. An IOSp$(3,1|4)$ transformation in superspace has the infinitesimal form
\begin{equation} \label{iosp}
\begin{split}
{Z'}^\La& =Z^\La+\,^\La\!\La_\Pi\,Z^\Pi+\Xi^\La, \\
\text{where} \quad _\La\La_\Pi =-&(-1)^{\La+\Pi+\La\Pi}\;_\Pi\La_\La \quad\text{(antisupersymmetric)},
\end{split}
\end{equation} 
with constant parameters $^\La\!\La_\Pi$ and $\Xi^\La$. Gauge supersymmetry, the analogue in superspace of general relativity, can be constructed by gauging IOSp$(3,1|4)$ on superspace in complete analogy to the manner (discussed in~\cite{gri}, for example) in which general relativity is obtained by gauging the Poincar\'{e} group on space-time.

Consideration of the geometry of a (4,4)-dimensional supermanifold thus leads naturally to the group IOSp$(3,1|4)$; however this group has nothing to do with physics. The group relevant to supersymmetric physics is the super Poincar\'{e} group, which gives the following infinitesimal transformation on superspace~\cite{wes3,dew}:
\begin{eqnarray}
{x'}^\mu\!\!&=&\!\!x^\mu+a^\mu+\zeta^\mu_{\ \,\nu}\,x^\nu+\si^\mu_{\ \,\al\dot{\al}}(\xi^\al\,\te^{\dot{\al}}-\te^\al\,\xi^{\dot{\al}}), \label{spsusp1} \\[5pt]
{\te'}^\al\!\!&=&\!\!\te^\al+\xi^\al+\frac{1}{2}\zeta_{\mu\nu}\,\si^{\mu\nu\al}_{\ \ \ \;\be}\,\te^\be, \label{spsusp2} \\[7pt]
{\te'}^{\dot{\al}}\!\!&=&\!\!\te^{\dot{\al}}+\xi^{\dot{\al}}+\frac{1}{2}\zeta_{\mu\nu}\,\si^{\mu\nu\dot{\al}}_{\ \ \ \;\dot{\be}}\,\te^{\dot{\be}}. \label{spsusp3}
\end{eqnarray}
Note that the absence of a dimensionful constant in the last term of (\ref{spsusp1}) is a consequence of (\ref{Coord}); this term is also remarkable because in it we have the translation parameter of $\te$ appearing as a ``rotation'' parameter of $x$. The metric (\ref{Met}) is, of course, not invariant under a super Poincar\'{e} transformation; this becomes obvious when we write (\ref{spsusp1})--(\ref{spsusp3}) in matrix form: \vspace{2mm}
\begin{equation} \label{spmat}
\left(\begin{array}{c} {x'}^\mu \\ {\te'}^\al \\ {\te'}^{\dot{\al}} \end{array}\right)
=\left(\begin{array}{c} x^\mu \\ \te^\al \\ \te^{\dot{\al}} \end{array}\right)+
\left(\begin{array}{ccc} \la^\mu_{\ \nu} & \si^\mu_{\ \,\be\dot{\al}}\;\xi^{\dot{\al}} & \si^\mu_{\ \,\al\dot{\be}}\;\xi^{\al} \\
0 & \frac{1}{2}\la_{\mu\nu}\;\si^{\mu\nu\al}_{\ \ \ \;\be} & 0 \\
0 & 0 & \frac{1}{2}\la_{\mu\nu}\;\si^{\mu\nu\dot{\al}}_{\ \ \ \;\dot{\be}} \end{array}\right)
\left(\begin{array}{c} x^\nu \\ \te^\be \\ \te^{\dot{\be}} \end{array}\right) +
\left(\begin{array}{c} a^\mu \\ \xi^\al \\ \xi^{\dot{\al}} \end{array}\right). \vspace{2mm}
\end{equation}
In the notation of (\ref{iosp}) the ``rotation'' matrix in (\ref{spmat}) is
\begin{gather} \vspace{5mm}
^\La\!\La_\Pi =\left(\begin{array}{ccc} \la^\mu_{\ \nu} & \si^\mu_{\ \,\be\dot{\al}}\;\xi^{\dot{\al}} & \si^\mu_{\ \,\al\dot{\be}}\;\xi^{\al} \\
0 & \frac{1}{2}\la_{\mu\nu}\;\si^{\mu\nu\al}_{\ \ \ \;\be} & 0 \\
0 & 0 & \frac{1}{2}\la_{\mu\nu}\;\si^{\mu\nu\dot{\al}}_{\ \ \ \;\dot{\be}} \end{array}\right)  \\[8pt]
\ \Rightarrow\quad _\La\La_\Pi=\,_\La H_\Si\;^\Si\!\La_\Pi=\left(\begin{array}{ccc} \la_{\mu\nu} & \si_{\mu\be\dot{\al}}\;\xi^{\dot{\al}} & \si_{\mu\al\dot{\be}}\;\xi^{\al} \\
0 & -\frac{1}{2}k^2\la_{\mu\nu}\;\si^{\mu\nu}_{\ \ \;\al\be} & 0 \\
0 & 0 & \frac{1}{2}k^2\la_{\mu\nu}\;\si^{\mu\nu}_{\ \ \;\dot{\al}\dot{\be}} \end{array}\right)  \label{rotsp} 
\end{gather} 
and the latter is clearly not antisupersymmetric; hence (\ref{spmat}) does not leave (\ref{Met}) unchanged. One can proceed to find geometrical objects such as a metric and a connection that {\it are} invariant under the super Poincar\'{e} group of transformations (\ref{spmat}), as is done in~\cite{dew} and~\cite{wes3}. The approach here however will be to explore the relationship between the super Poincar\'{e} group and the ``natural'', canonical structures on superspace.

We can construct an antisupersymmetric matrix from (\ref{rotsp}) by inserting additional elements; the least modification necessary to achieve this results in the matrix
\begin{gather} \vspace{5mm}
_\La\La_\Pi =\left(\begin{array}{ccc} \la_{\mu\nu} & \si_{\mu\be\dot{\al}}\;\xi^{\dot{\al}} & \si_{\mu\al\dot{\be}}\;\xi^{\al} \\
\si_{\nu\al\dot{\al}}\;\xi^{\dot{\al}} & -\frac{1}{2}k^2\la_{\mu\nu}\;\si^{\mu\nu}_{\ \ \;\al\be} & 0 \\
\si_{\nu\al\dot{\al}}\;\xi^\al & 0 & \frac{1}{2}k^2\la_{\mu\nu}\;\si^{\mu\nu}_{\ \ \;\dot{\al}\dot{\be}} \end{array}\right)  \\[8pt]
\ \Rightarrow\quad ^\La\!\La_\Pi=\left(\begin{array}{ccc} \la^\mu_{\ \nu} & \si^\mu_{\ \,\be\dot{\al}}\;\xi^{\dot{\al}} & \si^\mu_{\ \,\al\dot{\be}}\;\xi^{\al} \\
-\frac{1}{k^2}\si_{\nu\ \dot{\al}}^{\ \al}\;\xi^{\dot{\al}} & \frac{1}{2}\la_{\mu\nu}\;\si^{\mu\nu\al}_{\ \ \ \;\be} & 0 \\
\frac{1}{k^2}\si_{\nu\al}^{\ \ \;\dot{\al}}\;\xi^\al & 0 & \frac{1}{2}\la_{\mu\nu}\;\si^{\mu\nu\dot{\al}}_{\ \ \ \;\dot{\be}} \end{array}\right). \label{modrotsp}
\end{gather} 
By replacing the ``rotation'' matrix in (\ref{spmat}) with (\ref{modrotsp}) we obtain a transformation that does leave the metric (\ref{Met}) unchanged: 
\begin{equation} \label{modspmat}
\left(\begin{array}{c} {x'}^\mu \\ {\te'}^\al \\ {\te'}^{\dot{\al}} \end{array}\right)
=\left(\begin{array}{c} x^\mu \\ \te^\al \\ \te^{\dot{\al}} \end{array}\right)+
\left(\begin{array}{ccc} \la^\mu_{\ \nu} & \si^\mu_{\ \,\be\dot{\al}}\;\xi^{\dot{\al}} & \si^\mu_{\ \,\al\dot{\be}}\;\xi^{\al} \\
-\frac{1}{k^2}\si_{\nu\ \dot{\al}}^{\ \al}\;\xi^{\dot{\al}} & \frac{1}{2}\la_{\mu\nu}\;\si^{\mu\nu\al}_{\ \ \ \;\be} & 0 \\
\frac{1}{k^2}\si_{\nu\al}^{\ \ \;\dot{\al}}\;\xi^\al & 0 & \frac{1}{2}\la_{\mu\nu}\;\si^{\mu\nu\dot{\al}}_{\ \ \ \;\dot{\be}} \end{array}\right)
\left(\begin{array}{c} x^\nu \\ \te^\be \\ \te^{\dot{\be}} \end{array}\right) +
\left(\begin{array}{c} a^\mu \\ \xi^\al \\ \xi^{\dot{\al}} \end{array}\right).
 \end{equation}
The transformations (\ref{modspmat}) are a subset of the IOSp$(3,1|4)$ transformations (\ref{iosp}), but they do not form a group (as is verified by a tedious calculation). Thus, the transformations (\ref{modspmat}) are a subset of IOSp$(3,1|4)$, not a subgroup. Nevertheless the $k\to\infty$ limit of (\ref{modspmat}) does give a group---the group of super Poincar\'{e} transformations (\ref{spmat}). To reiterate this point, in the limit $k\to\infty$ the transformations (\ref{modspmat}) are no longer a subset of IOSp$(3,1|4)$, but rather form a new group, the super Poincar\'{e} group.

These considerations suggest how supergravity may be related to IOSp$(3,1|4)$ gauge theory. The latter gives us super one-form gauge potentials $A^A_{\ \,B\La}$ and $^A\!E_\La$ (the latter chosen to be the vielbein\footnote{We thus have an {\it affine connection}~\cite{kob} on the principle bundle with fibre IOSp$(3,1|4)$.}) with values in the super Lie algebra of IOSp$(3,1|4)$. Now we have seen how to obtain the infinitesimal super Poincar\'{e} group from infinitesimal IOSp$(3,1|4)$---extract all elements of infinitesimal  IOSp$(3,1|4)$ of the form (\ref{modspmat}) and take $k\to\infty$---so we can perform a similar operation with the respective Lie algebras. In this manner the potentials $A^A_{\ \,B\La}$ and $^A\!E_\La$ are turned into super Poincar\'{e}-algebra-valued objects and it is at this point that one might expect to see some physics.

Note that when $k\to\infty$ we lose the metric (\ref{Met}), though we can preserve a metric $\eta_{\mu\nu}$ in the bosonic sector. This is reminiscent of Cartan's space-time formulation of Newtonian gravity~\cite{mis}, wherein one also does not have a metric in the full space (space-time) but only in a subspace (3-space). This analogy will be pursued further below. 

\section{Super Lie algebra of IOSp$\mathbf{(3,1|4)}$ and the super Poincar\'{e} algebra} \label{iospsp}
In order to proceed we require the super Lie algebra of IOSp$(3,1|4)$, which we will derive by a method used in~\cite{wei} to obtain the Poincar\'{e} algebra. We shall then demonstrate explicitly that this super Lie algebra becomes the super Poincar\'{e} algebra when we select the elements corresponding to (\ref{modspmat}) and take the limit $k\to\infty$.

It will be convenient to write both indices of the OSp$(3,1|4)$ parameters $^\La\!\La_\Pi$ on the left; the rule for moving the lower index to the left is the same as if the upper index were absent~\cite{dew}:
\begin{equation} \label{Lashift}
^\La_{\ \,\Pi}\La:=(-1)^\Pi\;^\La\!\La_\Pi.
\end{equation}
We write the infinitesimal IOSp$(3,1|4)$ element as
\begin{equation} \label{infiosp}
G(1+\La,\Xi)=1+\frac{i}{2}J^\La_{\ \;\Pi}\;^\Pi_{\ \,\La}\La-iP_\La\;\Xi^\La.
\end{equation}
Eqn.~(\ref{infiosp}) does not serve to define the generatos $J^\La_{\ \;\Pi}$ and $P_\La$ completely; we must specify how the group element acts on a representation space. We define the element (\ref{infiosp}) to act on a pure vector $X$ in the representation space according to
\[
^i\!X'=\,^i\!X+\frac{i}{2}(-1)^{X(\La+\Pi)}\;^i\!\left[J^\La_{\ \;\Pi}(X)\right]\,^\Pi_{\ \,\La}\La-i(-1)^{X\La}\;^i\!\left[P_\La(X)\right]\Xi^\La.
\]
We arrange matters in this way so as to avoid the appearance in the group element (\ref{infiosp}) of factors of $(-1)$ that are dependent on representation-space indices, and so that the generators obey the super Lie algebra of the group.\footnote{These complicated issues are discussed both in generality and with many examples (including OSp$(m|n)$, but not IOSp$(m|n)$) in Chapters~3 and~4 of deWitt~\cite{dew}}

We denote a non-infinitesimal group element by $G(L,A)$ and consider the product
\begin{equation} \label{GGG-1}
G(L,A)G(1+\La,\Xi)G(L,A)^{-1}.
\end{equation}
Now a non-infinitesimal transformation $(L,A)$, given by
\[
{Z'}^\La =\,^\La\!L_\Pi\,Z^\Pi+A^\La,
\]
has as its inverse the transformation $(L^{-1},-L^{-1}A)$:
\[
{{Z'}'}^\La =\,^\La\!L^{-1}_{\ \ \;\Pi}\;{Z'}^\Pi-\,^\La\!L^{-1}_{\ \ \;\Pi}\;A^\Pi=Z^\La.
\]
Also, a transformation $(L,A)$ followed by a transformation $(\overline{L},\overline{A})$ is a transformation $(\overline{L}L,\overline{L}A+\overline{A})$. Hence the product (\ref{GGG-1}) is
\[
G(L,A)G(1+\La,\Xi)G(L^{-1},-L^{-1}A)=G(1+L\La L^{-1},-L\La L^{-1}A+L\Xi).
\]
To first order in $\La,\Xi$ we therefore have from (\ref{infiosp})
\begin{gather}
\hspace{-40mm} G(L,A)\left[1+\frac{i}{2}J^\La_{\ \;\Pi}\;^\Pi_{\ \,\La}\La-iP_\La\;\Xi^\La\right]G(L,A)^{-1} \nonumber \\[7pt]
\hspace{20mm} =1+\frac{i}{2}J^\La_{\ \;\Pi}\;^\Pi_{\ \,\La}(L\La L^{-1})-iP_\La\;^\La(L\Xi-L\La L^{-1}A) \nonumber \\[5pt]
 [(\ref{Lashift})]=1+\frac{i}{2}(-1)^\La\;J^\La_{\ \;\Pi}\;^\Pi\!L_\Xi\;^\Xi\!\La_\Si\;^\Si\!L^{-1}_{\ \ \;\La}-iP
_\La\left(^\La\!L_\Pi\;\Xi^\Pi-\,^\La\!L_\Pi\;^\Pi\!\La_\Xi\;^\Xi\!L^{-1}_{\ \ \;\Si}\;A^\Si\right) \label{GGG-12}
\end{gather}

We wish to equate coefficients of $^\Pi_{\ \,\La}\La$ and $\Xi^\La$ in (\ref{GGG-12}). In doing this we must take account of the antisupersymmetry of $_\Pi\La_\La$, as expressed in (\ref{iosp}); from the antisupersymmetry we obtain
\begin{gather}
^\La\!\La_\Pi=-(-1)^{\Pi(\La+1)}\;_\Pi\La^\La \nonumber \\
\Longrightarrow \qquad ^\La_{\ \,\Pi}\La=-(-1)^{\Pi\Xi}\ ^\La\!H^\Xi\;_\Pi H_\Si\;^\Si_{\ \,\Xi}\La \nonumber \\[5pt]
\Longrightarrow \qquad ^\La_{\ \,\Pi}\La=\frac{1}{2}\left(^\La_{\ \,\Pi}\La-(-1)^{\Pi\Xi}\ ^\La\!H^\Xi\;_\Pi H_\Si\;^\Si_{\ \,\Xi}\La \right). \label{Laanti} 
\end{gather}
Using (\ref{Laanti}) and the fact that $_\La H_\Pi$ vanishes if $\scriptstyle\La$ and $\scriptstyle\Pi$ are of opposite type, we obtain from equating coefficients of  $^\Pi_{\ \,\La}\La$  in (\ref{GGG-12})
\begin{equation} \label{GJG}
\begin{split}
G(L,A)J^\La_{\ \;\Pi}\,G(L,A)^{-1}=&(-1)^{\Si+\Pi(\La+\Si)+\La\Si}\;J^\Si_{\ \;\Xi}\;^\Xi\!L_\Pi\;^\La\!L^{-1}_{\ \ \;\Si} \\
+(-1)^{\Pi\La}\;&P_\Xi\;^\Xi\!L_\Pi\;^\La\!L^{-1}_{\ \ \;\Si}\;A^\Si-P_\Xi\;^\Xi\!L_\Phi\;^\Phi\!H^\La\;_\Up H_\Pi\;^\Up\!L^{-1}_{\ \ \;\Si}\;A^\Si,
\end{split}
\end{equation}
while equating coefficients of $\Xi^\La$ in (\ref{GGG-12}) gives
\begin{equation} \label{GPG}
G(L,A)P_\La\,G(L,A)^{-1}=P_\Pi\;^\Pi\!L_\La.
\end{equation}

We now apply (\ref{GJG}) and (\ref{GPG}) to the case where $G(L,A)$ is an infinitesimal element $G(1+\La,\Xi)$; using (\ref{infiosp}), (\ref{GJG}) now gives for terms of order $\La$ and $\Xi$ \vspace{2mm}
\begin{equation} \label{JPJcom}
\begin{split}
i\left[\frac{1}{2}J^\Si_{\ \;\Xi}\;\,^\Xi_{\ \,\Si}\La-P_\Si\;^\Si\Xi,J^\La_{\ \;\Pi}\right]=&\,J^\La_{\ \;\Xi}\;^\Xi\!\La_\Pi-(-1)^{\Si+\Pi(\La+\Si)+\La\Si}\;J^\Si_{\ \;\Pi}\;^\La\!\La_\Si  \\
&+(-1)^{\Pi\La}\;P_\Pi\;\Xi^\La-P_\Xi\;^\Xi\!H^\La\;_\Si H_\Pi\;\Xi^\Si,
\end{split}
\end{equation}
while (\ref{GPG}) gives
\begin{equation} \label{JPPcom}
i\left[\frac{1}{2}J^\Pi_{\ \;\Si}\;\,^\Si_{\ \,\Pi}\La-P_\Pi\;^\Pi\Xi,P_\La\right]=P_\Pi\;^\Pi\!\La_\La. \vspace{2mm}
\end{equation}
Equating coefficients of $^\Xi_{\ \,\Si}\La$ and $\Xi^\Si$ in (\ref{JPJcom}) and (\ref{JPPcom}) we obtain
\begin{eqnarray}
J^\La_{\ \;\Pi}\;J^\Si_{\ \;\Xi}-(-1)^{(\La+\Pi)(\Si+\Xi)}\;J^\Si_{\ \;\Xi}\;J^\La_{\ \;\Pi}\!\!\!&=&\!\!\!i\left[(-1)^\Si\;^\Si\!\de_\Pi\;J^\La_{\ \;\Xi}-(-1)^{\Pi(1+\Si)}\;_\Pi H_\Xi\;^\Up\!H^\Si\;J^\La_{\ \;\Up}\right. \nonumber \\[4pt]
& &\hspace{-40mm}\left.-(-1)^{\Pi(\La+\Si)+\La\Si}\;^\La\!\de_\Xi\;J^\Si_{\ \;\Pi}+(-1)^{\Pi(\La+\Up)}\;^\La\!H^\Si\;_\Up H_\Xi\;J^\Up_{\ \;\Pi}\right], \label{JJcom} \\[4pt]
J^\La_{\ \;\Pi}\;P_\Si-(-1)^{\Si(\La+\Pi)}\;P_\Si\;J^\La_{\ \;\Pi}\!\!\!&=&\!\!\!i\left[-(-1)^{\Pi\La}\;^\La\!\de_\Si\;P_\Pi+\,_\Si H_\Pi\;^\Xi\!H^\La\;P_\Xi\right], \label{JPcom} \\
P_\La\;P_\Pi-(-1)^{\La\Pi}\;P_\Pi\;P_\La\!\!\!&=&\!\!\!0. \label{PPcom}
\end{eqnarray}
The supercommutation relations (\ref{JJcom})--(\ref{PPcom}) constitute the super Lie algebra of \linebreak IOSp$(3,1|4)$.

Now consider the subset  (\ref{modspmat}) of IOSp$(3,1|4)$ transformations. In this subset the group parameters $^\La\!\La_\Pi$ and $\Xi^\La$ have the restrictions \vspace{2mm}
\begin{gather} 
^\al\!\La_\be=\frac{1}{2}\,_\mu\La_\nu\;\si^{\mu\nu\al}_{\ \ \ \;\be}, \qquad ^{\dot{\al}}\!\La_{\dot{\be}}=\frac{1}{2}\,_\mu\La_\nu\;\si^{\mu\nu\dot{\al}}_{\ \ \ \;\dot{\be}}, \qquad ^{\al}\!\La_{\dot{\al}}=0, \qquad ^{\dot{\al}}\!\La_{\al}=0, \label{modpar1} \\[4pt]
^\mu\!\La_\al=\si^\mu_{\ \,\al\dot{\al}}\;\Xi^{\dot{\al}}, \quad ^\mu\!\La_{\dot{\al}}=\si^\mu_{\ \,\al\dot{\al}}\;\Xi^{\al}, \quad ^\al\!\La_\mu=-\frac{1}{k^2}\si_{\mu\ \dot{\al}}^{\ \al}\;\Xi^{\dot{\al}}, \quad ^{\dot{\al}}\!\La_\mu=\frac{1}{k^2}\si_{\mu\al}^{\ \ \,\dot{\al}}\;\Xi^\al. \label{modpar2}
\end{gather}
Thus the only independent parameters in (\ref{modspmat}) are 
\begin{equation} \label{sppars0}
^\mu\!\La_\nu\ \ \text{and}\ \ \Xi^\La.
\end{equation}
We see from (\ref{infiosp}) that with the constraints (\ref{modpar1})--(\ref{modpar2}) the generators congugate to $^\nu_{\ \mu}\La$ are
\begin{equation} \label{modlagen}
J^\mu_{\ \,\nu}-\frac{1}{2}J^\al_{\ \,\be}\,\si_{\nu\ \ \,\al}^{\ \mu\be}-\frac{1}{2}J^{\dot{\al}}_{\ \,\dot{\be}}\,\si_{\nu\ \ \,\dot{\al}}^{\ \mu\dot{\be}}.
\end{equation}
The generators conjugate to $\Xi^\mu$ are still $P_\mu$, whereas the generators conjugate to $\Xi^\al$ are now\footnote{Recall that indices are raised and lowered by $^\La\!H^\Pi$ and $_\La H_\Pi$, so that $J^\mu_{\ \,\dot{\al}}=-J^{\ \mu}_{\dot{\al}}=-J^{\La\mu}\;_\La H_{\dot{\al}}=k^2J^{\dot{\be}\mu}\,\ep_{\dot{\be}\dot{\al}}$; but the definition of the Infeld-van der Waerden symbols requires us to raise and lower the spinor indices of these objects with $\ep\,$s, e.g.\ $\si^\mu_{\ \,\al\dot{\be}}=-\si_{\mu\al}^{\ \ \;\dot{\al}}\,\ep_{\dot{\be}\dot{\al}}$.}
\begin{equation} \label{modxigen1}
P_\al+\frac{1}{2}J^{\dot{\al}}_{\ \,\mu}\,\si^\mu_{\ \,\al\dot{\al}}-\frac{1}{2k^2}J^\mu_{\ \,\dot{\al}}\,\si_{\mu\al}^{\ \ \,\dot{\al}}=P_\al+J^{\dot{\al}}_{\ \,\mu}\,\si^\mu_{\ \,\al\dot{\al}},
\end{equation}
and those conjugate to $\Xi^{\dot{\al}}$ are
\begin{equation} \label{modxigen2}
P_{\dot{\al}}+\frac{1}{2}J^{\al}_{\ \,\mu}\,\si^\mu_{\ \,\al\dot{\al}}-\frac{1}{2k^2}J^\mu_{\ \,\al}\,\si_{\mu\ \dot{\al}}^{\ \al}=P_{\dot{\al}}+J^{\al}_{\ \,\mu}\,\si^\mu_{\ \,\al\dot{\al}}.
\end{equation}

We now show that, in the limit $k\to\infty$, (\ref{sppars0})--(\ref{modxigen2}) (plus $P_\mu$) are the parameters and generators of the super Poincar\'{e} group. It might appear that we obtain different generators depending on whether we take the $k\to\infty$ limit of the left-hand sides or right-hand sides of (\ref{modxigen1}) and (\ref{modxigen2}), but this is not the case: the two $J$-terms on the left-hand sides have a factor of one half because each represents the same independent generator of OSp$(3,1|4)$ which is double counted in the summation in (\ref{infiosp}); if we take $k\to\infty$ on the left-hand sides then we must drop the factor of one half in the remaining $J$-term since it is then no longer double counted. Thus the right-hand sides of  (\ref{modxigen1}) and (\ref{modxigen2}) represent the correct $k\to\infty$ limit of these generators.

The $k\to\infty$ limit of the subset (\ref{modspmat}) of IOSp$(3,1|4)$ transformations thus has group parameters 
\begin{equation} \label{spparsder}
^\mu\!\La_\nu=:\la^\mu_{\ \,\nu}, \qquad \Xi^\al=:\xi^\al, \qquad \Xi^{\dot{\al}}=:\xi^{\dot{\al}}, \qquad \text{and} \quad \Xi^\mu=:a^\mu
\end{equation}
with, respectively, group generators
\begin{gather}
J^\mu_{\ \,\nu}-\frac{1}{2}J^\al_{\ \,\be}\,\si_{\nu\ \ \,\al}^{\ \mu\be}-\frac{1}{2}J^{\dot{\al}}_{\ \,\dot{\be}}\,\si_{\nu\ \ \,\dot{\al}}^{\ \mu\dot{\be}}=:-\mathcal{J}^\mu_{\ \,\,\nu}, \label{spJder} \\
P_\al+J^{\dot{\al}}_{\ \,\mu}\,\si^\mu_{\ \,\al\dot{\al}}=:-iQ_\al, \label{spQder} \\
P_{\dot{\al}}+J^{\al}_{\ \,\mu}\,\si^\mu_{\ \,\al\dot{\al}}=:-iQ_{\dot{\al}}, \label{spQbarder} \\
\text{and} \quad P_\mu.
\end{gather}
It is now a straightforward, though tedious, matter to calculate the algebra of the new generators using (\ref{JJcom})--(\ref{PPcom}) and (\ref{Met}): from the definitions (\ref{spJder})--(\ref{spQbarder}) we obtain, in the limit $k\to\infty$,
\begin{eqnarray*}
[P_\mu,P_\nu]&\!\!\!=&\!\!\!0=[P_\mu,Q_\al]=[P_\mu,Q_{\dot{\al}}], \\
\{Q_\al,Q_\be\}&\!\!\!=&\!\!\!0=\{Q_{\dot{\al}},Q_{\dot{\be}}\}, \\
\{Q_\al,Q_{\dot{\al}}\}\!\!\!&=&\!\!\!2i\si^\mu_{\ \,\al\dot{\al}}\,P_\mu, \\
\left[Q_\al,\mathcal{J}^\mu_{\ \,\,\nu}\right]\!\!\!&=&\!\!\!\,i\si^{\mu\ \,\be}_{\ \,\nu\ \,\al}\,Q_\be, \\
\left[Q_{\dot{\al}},\mathcal{J}^\mu_{\ \,\,\nu}\right]\!\!\!&=&\!\!\!\,i\si^{\mu\ \,\dot{\be}}_{\ \,\nu\ \,\dot{\al}}\,Q_{\dot{\be}}, \\[4pt]
[\mathcal{J}^\mu_{\ \,\,\nu},\mathcal{J}^\la_{\ \,\,\rho}]\!\!\!&=&\!\!\!i(-\eta^\la_{\ \,\nu}\,\mathcal{J}^\mu_{\ \;\rho}+\eta_{\nu\rho}\,\mathcal{J}^{\mu\la}-\eta^\mu_{\ \,\rho}\,\mathcal{J}^{\ \la}_{\nu}+\eta^{\mu\la}\,\mathcal{J}_{\nu\rho}), \\[5pt]
[\mathcal{J}^\mu_{\ \,\,\nu},P_\la]\!\!\!&=&\!\!\!i(\eta^\mu_{\ \,\la}\,P_\nu-\eta_{\nu\la}\,P^\mu), \\
\left[P_\mu,P_\nu\right]\!\!\!&=&\!\!\! 0.
\end{eqnarray*}
This is the super Poincar\'{e} algebra, as expected. The infinitesimal group element is, from (\ref{infiosp}), (\ref{modpar1})--(\ref{modpar2}) and (\ref{spJder})--(\ref{spQbarder}),
\[
G(1+\la,\Xi)=1+\frac{i}{2}\mathcal{J}^\mu_{\ \,\,\nu}\,\la_\mu^{\ \,\nu}-iP_\mu\,a^\mu-Q_\al\,\xi^\al-Q_{\dot{\al}}\,\xi^{\dot{\al}}.
\]

\section{IOSp$\mathbf{(3,1|4)}$ gauge theory}
Equipped with the super Lie algebra (\ref{JJcom})--(\ref{PPcom}) we can now construct the gauge theory of IOSp$(3,1|4)$. The non-super analogue of this is Poincar\'{e} gauge theory, and the corresponding formulae of IOSp$(3,1|4)$ gauge theory will differ only through index-dependent factors of minus one.

We consider the super fibre bundle whose base space is superspace and whose typical fibre is IOSp$(3,1|4)$. An affine connection on this bundle has a pull-back $A$ which takes the form (see (\ref{infiosp}))
\begin{equation} \label{iosppot}
A=A_\La\;dZ^\La=\left(\frac{i}{2}(-1)^A\;J^A_{\ \;B}\;A^B_{\ \,A\La}-iP_A\;^A\!E_\La\right)dZ^\La,
\end{equation}
where $^A\!E_\La$ is the superspace vielbein which relates the coordinate frame to the orthosymplectic frame. Under an infinitesimal gauge transformation $A$ transforms to
\[
\begin{split}
A'=&\,G(1+\La,\Xi)^{-1}AG(1+\La,\Xi)+G(1+\La,\Xi)^{-1}\,dG(1+\La,\Xi)  \\[6pt]
[(\ref{infiosp})]\,=&\,A-\frac{i}{2}[J^A_{\ \;B},A]\;^B_{\ \,A}\La+i[P_A,A]\,\Xi^A+\frac{i}{2}J^A_{\ \;B}\;\,d\,^B_{\ \,A}\La-iP_A\,d\Xi^A.
\end{split}
\]
We use (\ref{iosppot}) and (\ref{JJcom})--(\ref{PPcom}) to evaluate the supercommutators and find
\[
\begin{split}
A'=&\,A+\frac{i}{2}J^A_{\ \;D}\;A^D_{\ \,B\La}\;dZ^\La\;^B_{\ \,A}\La-\frac{i}{2}(-1)^{C(1+A)+B(A+C)}\;J^C_{\ \;B}\;A^A_{\ \,C\La}\;dZ^\La\;^B_{\ \,A}\La \\[5pt]
&+i(-1)^{BA}\;P_B\;^A\!E_\La\;dZ^\La\;^B_{\ \,A}\La-iP_C\;A^C_{\ \,A\La}\;dZ^\La\;\Xi^A+\frac{i}{2}J^A_{\ \;B}\;\,d\,^B_{\ \,A}\La-iP_A\,d\Xi^A \\
=:&A+\de A.
\end{split}
\]
Comparing this with (\ref{iosppot}) we see that the component parts of the gauge potential have the gauge transformations
\begin{gather}
\de\,^A\!E_\La=\Xi^A_{\ \,,\La}+(-1)^{B(A+1)}\;\;\Xi^B\;A^A_{\ \,B\La}-\,^A\!\La_B\;^B\!E_\La,  \label{Etrans} \\
\de\,A^A_{\ \,B\La}=\,^A\!\La_{B,\La}+(-1)^{(C+B)(A+C)}\;\;^C\!\La_B\;A^A_{\ \,C\La}-\,^A\!\La_C\;A^C_{\ \,B\La}, \label{Atrans}
\end{gather}
The commutator of two small gauge transformations $\de(\La_1,\Xi_1)$ and $\de(\La_2,\Xi_2)$ of $A$ reads, to leading order,
\[
\left[\de(\La_1,\Xi_1),\de(\La_2,\Xi_2)\right]A=\de(\La',\Xi')A,
\]
where
\[
\begin{split}
^A\!{\La'}_B&=\,^A\!\La_{1C}\;^C\!\La_{2B}-\,^A\!\La_{2C}\;^C\!\La_{1B}, \\
{\Xi'}^A&=\,^A\!\La_{1B}\;\Xi^B_2-\,^A\!\La_{2B}\;\Xi^B_1.
\end{split}
\]

The pull-back of the gauge curvature is
\[
\begin{split}
R=&\frac{1}{2}(-1)^{\La\Pi}\;R_{\La\Pi}\;dZ^\La\wedge dZ^\Pi \\[5pt]
=&\frac{1}{2}(A_{\Pi,\La}-(-1)^{\La\Pi}A_{\La,\Pi}+(-1)^{\La\Pi}A_\La\,A_\Pi-A_\Pi\,A_\La)dZ^\La\wedge dZ^\Pi \\[7pt]
=&\left(\frac{i}{4}(-1)^{B+\La\Pi}\;J^B_{\ \;A}\;R^A_{\ \,B\La\Pi}-\frac{i}{2}(-1)^{\La\Pi}\;P_A\;R^A_{\ \,\La\Pi}\right)dZ^\La\wedge dZ^\Pi,
\end{split}
\]
where
\begin{eqnarray}
 R^A_{\ \,\La\Pi}=& &\hspace{-10mm}(-1)^{\La\Pi}\;\;^A\!E_{\Pi,\La}-\,^A\!E_{\La,\Pi}+(-1)^{\La B}\;A^A_{\ \,B\La}\;^B\!E_\Pi-(-1)^{\Pi(B+\La)}\;A^A_{\ \,B\Pi}\;^B\!E_\La,\ \ \ \ \ \ \ \  \label{Tor} \\
R^A_{\ \,B\La\Pi}\!\!\!&=&\!\!\!(-1)^{\La\Pi}\; A^A_{\ \,B\Pi,\La}-A^A_{\ \,B\La,\Pi}+(-1)^{\La(C+B)}\; A^A_{\ \,C\La}\,A^C_{\ \,B\Pi} \nonumber \\
& &-(-1)^{\Pi(C+B+\La)}\; A^A_{\ \,C\Pi}\,A^C_{\ \,B\La}. \label{Cur}
\end{eqnarray}
We recognise $R^A_{\ \,\La\Pi}$ as the torsion supertensor and $R^A_{\ \,B\La\Pi}$ as the Riemann supertensor:
\begin{gather}
R^A_{\ \,BC}=(-1)^{(\La+B)\Pi}\;R^A_{\ \,\La\Pi}\;^\La\!E_B\;^\Pi\!E_C, \\
R^A_{\ \,BCD}=(-1)^{(\La+C)\Pi}\;R^A_{\ \,B\La\Pi}\;^\La\!E_C\;^\Pi\!E_D.
\end{gather}
The gauge curvatures satisfy the following Bianchi identities:
\begin{gather}
R^A_{\ \,\La\Pi|\Si}+(-1)^{\La(\Pi+\Si)}\;R^A_{\ \,\Pi\Si|\La}+(-1)^{\Si(\La+\Pi)}\;R^A_{\ \,\Si\La|\Pi}=(-1)^{(B+\La)(\Pi+\Si)}\;R^A_{\ \,B\Pi\Si}\,^B\!E_\La \nonumber \\
+(-1)^{B(\Si+\La)+\Si(\La+\Pi)}\;R^A_{\ \,B\Si\La}\,^B\!E_\Pi+(-1)^{B(\La+\Pi)}\;R^A_{\ \,B\La\Pi}\,^B\!E_\Si, \label{Bian1}  \\
R^A_{\ \,B\La\Pi|\Si}+(-1)^{\La(\Pi+\Si)}\;R^A_{\ \,B\Pi\Si|\La}+(-1)^{\Si(\La+\Pi)}\;R^A_{\ \,B\Si\La|\Pi}=0,  \label{Bian2}
\end{gather}
where $\scriptstyle |$ denotes a covariant derivative with connection $A^A_{\ \,B\La}$ that acts only on ortho\-sym\-plectic-frame indices.

We could now go on to construct gauge supersymmetry, the superspace analogue of general relativity. The procedure corresponds exactly to the formulation of general relativity as Poincar\'{e} gauge theory. One takes the superspace version of the Hilbert action and finds that it is gauge invariant if and only if the torsion supertensor vanishes, which is in turn the requirement that the OSp$(3,1|4)$ potential $A^A_{\ \,B\La}$ satisfy its equation of motion. However, as discussed in the next section, the dynamics of gauge supersymmetry has no relevance for our derivation of supergravity. All of the details of IOSp$(3,1|4)$ gauge theory necessary to obtain supergravity have been presented so we now turn to the derivation.

\section{Supergravity}
Section~\ref{iospsp} showed how to obtain the super Poincar\'{e} algebra from the super Lie algebra of IOSp$(3,1|4)$. We now use this method to turn the gauge potential and curvature of IOSp$(3,1|4)$ gauge theory into super-Poincar\'{e}-algebra valued quantities. The result is a blend of gauge theory and the super Poincar\'{e} group, both of which are highly significant from a physical standpoint, so we may hope to obtain some supersymmetric physics. 

In curved superspace the canonical metric (\ref{Met}) with coordinates (\ref{Coord}) cannot, in general, be introduced in an extended region. The corresponding objects in curved superspace are the orthosymplectic frame $E_A$ and the orthosymplectic-frame components of the metric, which we choose in line with (\ref{Coord}) and (\ref{Met}):
\begin{gather}
[E_m]=1, \qquad [E_a]=[E_{\dot{a}}]=\sfrac{1}{2}, \label{Eorth} \\[8pt]
_A H_B=\left(\begin{array}{ccc} \eta_{mn} & 0 & 0 \\
0 & k^2\ep_{ab} & 0 \\
0 & 0 & -k^2\ep_{\dot{a}\dot{b}} \end{array}\right) \quad
\Rightarrow \quad ^A H^B=\left(\begin{array}{ccc} \eta^{mn} & 0 & 0 \\
0 & -\frac{1}{k^2}\ep^{ab} & 0 \\
0 & 0 & \frac{1}{k^2}\ep^{\dot{a}\dot{b}} \end{array}\right). \label{Horth}
\end{gather}
We choose the superspace coordinates $Z^\La$ to all have the same, standard, units
\begin{equation} \label{Zcoord}
[Z^\La]=-1\quad \forall \quad {\scriptstyle \La} 
\end{equation}
so that the metric coordinate components 
\[
_\La G_\Pi=\,_\La E^A\;_AH_B\;^B\!E_\Pi
\]
are dimensionless. In flat superspace the frame $E_A$ may also be taken as a coordinate frame and the simplest choice of vielbein (now a coordinate transformation) is then, in view of (\ref{Eorth})--(\ref{Zcoord}),
\begin{equation} \label{Eflat}
^A\!E_\La=\left(\begin{array}{ccc} ^m\de_\mu & 0 & 0 \\
0 & \frac{1}{k}\,^a\de_\al & 0 \\
0 & 0 & \frac{1}{k}\,^{\dot{a}}\de_{\dot{\al}} \end{array}\right).
\end{equation}
We therefore expect to be able to choose a vielbein that reduces to (\ref{Eflat}) when superspace is flat.

Following the path we have set out, we must extract the part of the IOSp$(3,1|4)$ gauge potential corresponding to the infinitesimal transformations (\ref{modspmat}) and take $k\to\infty$. Since we wish to discover what happens to the IOSp$(3,1|4)$ gauge transformations in this super-Poincar\'{e} limit, we must perform the same operation on the parameters of infinitesimal gauge transformations. Thus, from (\ref{modspmat}), the translation potentials $^A\!E_\La$ and parameters $\Xi^A$ are to remain independent, whereas we impose the following constraints on the OSp$(3,1|4)$  potentials $A^A_{\ \,B\La}$ and parameters $^A\!\La_B$:
\begin{gather}
A^A_{\ B\La}=\left(\begin{array}{ccc} A^m_{\ \;n\La} & \si^m_{\ \ b\dot{a}}\;^{\dot{a}}\!E_\La & \si^m_{\ \ a\dot{b}}\;^{a}\!E_\La \\
-\frac{1}{k^2}\si_{n\ \,\dot{a}}^{\ \,a}\;^{\dot{a}}\!E_\La & \frac{1}{2}A^m_{\ \;n\La}\;\si_{m\ \;\;b}^{\ \;na} & 0 \\
\frac{1}{k^2}\si_{na}^{\ \ \,\dot{a}}\;^a\!E_\La & 0 & \frac{1}{2}A^m_{\ \;n\La}\;\si_{m\ \;\;\dot{b}}^{\ \;n\dot{a}} \end{array}\right),   \label{Con1}    \\[10pt]
^A\!\La_B=\left(\begin{array}{ccc} ^m\!\La_{\!\!\!\ n} & \si^m_{\ \,b\dot{a}}\;\Xi^{\dot{a}} & \si^m_{\ \,a\dot{b}}\;\Xi^{a} \\ 
-\frac{1}{k^2}\si_{n\ \,\dot{a}}^{\ \,a}\;\Xi^{\dot{a}} & \frac{1}{2}\,^m\!\La_n\;\si_{m\ \ b}^{\ \;na} & 0 \\
\frac{1}{k^2}\si_{na}^{\ \ \,\dot{a}}\;\Xi^a & 0 & \frac{1}{2}\,^m\!\La_n\;\si_{m\ \ \dot{b}}^{\ \;n\dot{a}} \end{array}\right). \label{Con2}
\end{gather}
Note that $^A\!\La_B$ and $\Xi^A$ are the parameters of a gauge transformation and are therefore functions of the superspace coordinates $Z^\La$. From the derivation in Section~\ref{iospsp} we know that, with the constraints (\ref{Con1}) and the limit $k\to\infty$, the gauge potential (\ref{iosppot}) is super-Poincar\'{e}-algebra valued. Imposing (\ref{Con1}) and (\ref{Con2}) on the gauge transformation (\ref{Etrans}) of $^A\!E_\La$ gives
\begin{eqnarray}
\de\,^m\!E_\La&=&\Xi^m_{\ \;,\La}+\Xi^n\;A^m_{\ \;n\La}-2\si^m_{\ \ a\dot{a}}(\Xi^a\;^{\dot{a}}\!E_\La+\Xi^{\dot{a}}\;^a\!E_\La)-\,^m\!\La_n\;^n\!E_\La,  \label{Etrans1} \\[8pt]
\de\,^a\!E_\La&=&\Xi^a_{\ ,\La}+\frac{1}{2}\,\Xi^b\;A^m_{\ \;n\La}\;\si_{m\ \;\;b}^{\ \;na}-\frac{1}{k^2}\si_{m\ \,\dot{a}}^{\ \;\,a}\;\Xi^m\;^{\dot{a}}\!E_\La  \nonumber  \\[8pt]
& &-\frac{1}{2}\,^m\!\La_n\;\si_{m\ \;\;b}^{\ \;na}\;^b\!E_\La+\frac{1}{k^2}\si_{m\ \,\dot{a}}^{\ \;\,a}\;\Xi^{\dot{a}}\;^m\!E_\La, \label{Etrans2} \\[8pt]
\de\,^{\dot{a}}\!E_\La&=&\Xi^{\dot{a}}_{\ ,\La}+\frac{1}{2}\,\Xi^{\dot{b}}\;A^m_{\ \,n\La}\;\si_{m\ \ \dot{b}}^{\ \;n\dot{a}}+\frac{1}{k^2}\si_{ma}^{\ \ \ \dot{a}}\;\Xi^m\;^a\!E_\La \nonumber \\[8pt]
& &-\frac{1}{2}\,^m\!\La_n\;\si_{m\ \ \dot{b}}^{\ \;n\dot{a}}\;^{\dot{b}}E_\La-\frac{1}{k^2}\si_{ma}^{\ \ \ \dot{a}}\;\Xi^a\;^m\!E_\La. \label{Etrans3}
\end{eqnarray}
On the other hand, with the constraints (\ref{Con1}) the only independent $A^A_{\ B\La}$ are now the $6\times 8$ independent $A^m_{\ \,n\La}$. The gauge transformation of $A^m_{\ \,n\La}$ is obtained from (\ref{Atrans}) with (\ref{Con1}) and (\ref{Con2}):
\begin{eqnarray} 
\de\,A^m_{\ \,n\La}&=&\,^m\!\La_{n,\La}+\,^r\!\La_n\;A^m_{\ \,r\La}-\frac{1}{k^2}\si_{n\ \,\dot{a}}^{\ \,a}\;\Xi^{\dot{a}}\;\si^m_{\ \ a\dot{b}}\;^{\dot{b}}\!E_\La+\frac{1}{k^2}\si_{na}^{\ \ \,\dot{a}}\;\Xi^a\;\si_{mb\dot{a}}\;^b\!E_\La \nonumber \\[8pt]
 & &-\,^m\!\La_r\;A^r_{\ n\La}-\frac{1}{k^2}\si^m_{\ \ a\dot{a}}\;\Xi^{\dot{a}}\;\si_{n\ \,\dot{b}}^{\ \,a}\;^{\dot{b}}\!E_\La+\frac{1}{k^2}\si^m_{\ \ a\dot{a}}\;\Xi^a\;\si_{nb}^{\ \ \,\dot{a}}\;^b\!E_\La. \label{Atransa}
\end{eqnarray}

In light of (\ref{Eflat}), it is clear from (\ref{Etrans1})--(\ref{Etrans3}) that we may choose the order $Z^\al$ and $Z^{\dot{\al}}$ terms of $\Xi^A$ so that a $\Xi$-transformation of $^A\!E_\La$ leaves its $Z^\al$,$Z^{\dot{\al}}$-independent part in the form
\begin{equation} \label{Ecurv}
^A\!E_\La(Z^\al=Z^{\dot{\al}}=0)=\left(\begin{array}{ccc} e^m_{\ \,\mu} & 0 & 0 \\
\frac{1}{2}\phi^a_{\ \mu} & \frac{1}{k}\,^a\de_\al & 0 \\
\frac{1}{2}\phi^{\dot{a}}_{\ \mu} & 0 & \frac{1}{k}\,^{\dot{a}}\de_{\dot{\al}} \end{array}\right),
\end{equation}
where $e^m_{\ \,\mu}$, $\phi^a_{\ \mu}$ and $\phi^{\dot{a}}_{\ \mu}$ are functions of $Z^\mu$ only ($\phi^a_{\ \mu}$ and $\phi^{\dot{a}}_{\ \mu}$ $a$-type). Similarly, (\ref{Atransa}) shows that we  may choose the order $Z^\al$ and $Z^{\dot{\al}}$ terms of $^m\!\La_n$ so that a $\La$-transformation of $A^m_{\ \,n\La}$ results in
\begin{equation} \label{Aal=0}
A^m_{\ \,n\al}(Z^\al=Z^{\dot{\al}}=0)=A^m_{\ \,n\dot{\al}}(Z^\al=Z^{\dot{\al}}=0)=0.
\end{equation} 
The standard choices (\ref{Ecurv}) and (\ref{Aal=0}) amount to a partial gauge fixing.

We identify the bosonic sector of superspace with space-time. Note that this is the largest sector of superspace for which a metric can be retained in the $k\to\infty$ limit. We see this both from (\ref{Horth}) and (\ref{Ecurv}); the latter shows that  the matrix $^A\!E_\La$ does not have an inverse when $k\to\infty$ since its body doesn't~\cite{dew}, but we retain a tetrad $^m\!E_\mu$ in space-time. We must then take account of the experimental fact that physical fields such as the tetrad show no dependence on the $a$-type coordinates. Accordingly, we must take as the physical fields in space-time
\[
\left.^A\!E_\La\right|\ \ \text{and}\ \  \left.A^m_{\ \,n\La}\right|,
\]
where $\mid$ means ``$Z^\al=Z^{\dot{\al}}=0$ and $k\to\infty$''.\footnote{The standard procedure of ``gauge completion'', used in both the Nath--Arnowitt and Wess--Zumino formulations, can be used to give $^A\!E_\La$ and $A^m_{\ \,n\La}$ a complete expansion in the $a$-type coordinates when auxiliary fields are added to the theory.~\cite{nat2,bri,wes2}} 
Then with the partial gauge fixing (\ref{Ecurv}) and (\ref{Aal=0}) we get a physical field content of
\begin{gather}
\left.^A\!E_\La\right|=\left(\begin{array}{ccc} e^m_{\ \,\mu} & 0 & 0 \\
\frac{1}{2}\phi^a_{\ \mu} & 0 & 0 \\
\frac{1}{2}\phi^{\dot{a}}_{\ \mu} & 0 & 0 \end{array}\right), \label{physE} \\
\left.A^m_{\ \,n\mu}\right|=:\Ga^m_{\ \ n\mu}(Z^\mu),    \qquad \left.A^m_{\ \,n\al}\right|=\left.A^m_{\ \,n\dot{\al}}\right|=0. \label{physA=0}
\end{gather}
Note from (\ref{Etrans1})--(\ref{Atransa}) that the transformation of the physical fields is determined solely by
\begin{gather}
\left.\Xi^m\right|=:\ep^m, \quad \left.\Xi^a\right|=:-\xi^a, \quad \left.\Xi^{\dot{a}}\right|=:-\xi^{\dot{a}},  \label{Par1} \\
\left.^m\!\La_n\right|=:\la^m_{\ \,n}, \label{Par2}
\end{gather}
so that (\ref{Par1})--(\ref{Par2}) are the physically relevant gauge parameters. With (\ref{physE})--(\ref{Par2}), the $k\to\infty$ limit of the transformations (\ref{Etrans1})--(\ref{Atransa}) is
\begin{eqnarray}
\de e^m_{\ \,\mu}&=&\ep^m_{\ \, ,\mu}+\ep^n\,\Ga^m_{\ \;n\mu}+\si^m_{\ \ a\dot{a}}(\xi^a\,\phi^{\dot{a}}_{\ \mu}+\xi^{\dot{a}}\,\phi^a_{\ \mu})-\la^m_{\ \;n}\,e^n_{\ \mu}, \label{spsg1} \\[5pt]
\de\phi^a_{\ \mu}&=&-2D_\mu\,\xi^a-\frac{1}{2}\la^m_{\ \;n}\,\si_{m\ \ \,b}^{\ \;na}\,\phi^b_{\ \mu}, \label{spsg2} \\[5pt]
\de\phi^{\dot{a}}_{\ \mu}&=&-2D_\mu\,\xi^{\dot{a}}-\frac{1}{2}\la^m_{\ \;n}\,\si_{m\ \ \,\dot{b}}^{\ \;n\dot{a}}\,\phi^{\dot{b}}_{\ \mu}, \label{spsg3} \\[5pt]
\de\Ga^m_{\ \ n\mu}&=&\la^m_{\ \;n,\mu}+\la^r_{\ n}\,\Ga^m_{\ \ r\mu}-\la^m_{\ \;r}\,\Ga^r_{\ n\mu}, \label{spsg4}
\end{eqnarray}
where $D_\mu$ is the familiar space-time covariant derivative operator with connection $\Ga^m_{\ \ n\mu}$ that acts only on orthonormal-frame and spinor indices. Eqns.~(\ref{spsg1})--(\ref{spsg4}) are the super Poincar\'{e} gauge transformations of supergravity~\cite{cha}, with gravitino
\begin{equation} \label{phidef2}
\psi_\mu^{\ a}=\frac{1}{\ka}\phi^a_{\ \mu}, \qquad \ka=\sqrt{8\pi G}.
\end{equation}

Turning now to the gauge curvatures, we impose the constraints (\ref{Con1}) on the torsion (\ref{Tor}) and take $k\to\infty$, obtaining
\begin{eqnarray}
R^m_{\ \;\La\Pi}&=&(-1)^{\La\Pi}\;\;^m\!E_{\Pi,\La}-\,^m\!E_{\La,\Pi}+A^m_{\ \;n\La}\;^n\!E_\Pi-(-1)^{\Pi\La}\;A^m_{\ \;n\Pi}\;^n\!E_\La  \nonumber \\
& &+2(-1)^\La\;\si^m_{\ \ a\dot{a}}(^{\dot{a}}\!E_\La\;^a\!E_\Pi+\,^a\!E_\La\;^{\dot{a}}\!E_\Pi),   \label{Tora}  \\[5pt]
R^a_{\ \La\Pi}&=&(-1)^{\La\Pi}\;\;^a\!E_{\Pi,\La}-\,^a\!E_{\La,\Pi}+\frac{1}{2}(-1)^\La\;A^m_{\ \;n\La}\;\si_{m\ \;\;b}^{\ \;na}\;^b\!E_\Pi \nonumber \\[5pt]
& &-\frac{1}{2}(-1)^{\Pi(1+\La)}\;A^m_{\ \;n\Pi}\;\si_{m\ \;\;b}^{\ \;na}\;^b\!E_\La, \label{Torb} \\[6pt]
R^{\dot{a}}_{\ \La\Pi}&=&(-1)^{\La\Pi}\;\;^{\dot{a}}\!E_{\Pi,\La}-\,^{\dot{a}}\!E_{\La,\Pi}+\frac{1}{2}(-1)^\La\;A^m_{\ \;n\La}\;\si_{m\ \;\;\dot{b}}^{\ \;n\dot{a}}\;^{\dot{b}}\!E_\Pi \nonumber \\[5pt]
& &-\frac{1}{2}(-1)^{\Pi(1+\La)}\;A^m_{\ \;n\Pi}\;\si_{m\ \;\;\dot{b}}^{\ \;n\dot{a}}\;^{\dot{b}}\!E_\La, \label{Torc}
\end{eqnarray}
while the same operation on the Riemann supertensor (\ref{Cur}) produces
\begin{eqnarray}
R^m_{\ \;n\La\Pi}&\!\!\!\!=\!\!\!\!&(-1)^{\La\Pi}\;A^m_{\ \;n\Pi,\La}-A^m_{\ \;n\La,\Pi}+A^m_{\ \;r\La}\;A^r_{\ n\Pi}-(-1)^{\Pi\La}\;A^m_{\ \;r\Pi}\;A^r_{\ n\La}, \label{Cura} \\
R^m_{\ \;a\La\Pi}&\!\!\!\!=\!\!\!\!&(-1)^{\La\Pi}\;\si^m_{\ \ a\dot{a}}\;^{\dot{a}}\!E_{\Pi,\La}-\si^m_{\ \ a\dot{a}}\;^{\dot{a}}\!E_{\La,\Pi}+(-1)^\La\;A^m_{\ \;n\La}\;\si^n_{\ \,a\dot{a}}\;^{\dot{a}}\!E_\Pi \nonumber \\[5pt]
& &+\frac{1}{2}\si^m_{\ \ b\dot{a}}\;^{\dot{a}}\!E_\La\;A^n_{\ r\Pi}\;\si_{n\ \:\,a}^{\ rb}-(-1)^{\Pi(1+\La)}\;A^m_{\ \;n\Pi}\;\si^n_{\ a\dot{a}}\;^{\dot{a}}\!E_\La \nonumber \\[5pt]
& &-\frac{1}{2}(-1)^{\Pi\La}\;\si^m_{\ \ b\dot{a}}\;^{\dot{a}}\!E_\Pi\;A^n_{\ r\La}\;\si_{n\ \:\,a}^{\ rb}, \label{Curb} \\[6pt]
R^m_{\ \;\dot{a}\La\Pi}&\!\!\!\!=\!\!\!\!&(-1)^{\La\Pi}\;\si^m_{\ \ a\dot{a}}\;^{a}\!E_{\Pi,\La}-\si^m_{\ \ a\dot{a}}\;^{a}\!E_{\La,\Pi}+(-1)^\La\;A^m_{\ \;n\La}\;\si^n_{\ \,a\dot{a}}\;^{a}\!E_\Pi \nonumber \\[5pt]
& &+\frac{1}{2}\si^m_{\ \ a\dot{b}}\;^{a}\!E_\La\;A^n_{\ r\Pi}\;\si_{n\ \:\,\dot{a}}^{\ r\dot{b}}-(-1)^{\Pi(1+\La)}\;A^m_{\ \;n\Pi}\;\si^n_{\ a\dot{a}}\;^{a}\!E_\La \nonumber \\[5pt]
& &-\frac{1}{2}(-1)^{\Pi\La}\;\si^m_{\ \ a\dot{b}}\;^{a}\!E_\Pi\;A^n_{\ r\La}\;\si_{n\ \:\,\dot{a}}^{\ r\dot{b}}, \label{Curc} \\[6pt]
R^a_{\ b\La\Pi}&\!\!\!\!=\!\!\!\!&\frac{1}{2}(-1)^{\La\Pi}\;A^m_{\ \;n\Pi,\La}\;\si_{m\ \;\;b}^{\ \;na}-\frac{1}{2}(-1)^{\La\Pi}\;A^m_{\ \;n\La,\Pi}\;\si_{m\ \;\;b}^{\ \;na} \nonumber \\[5pt]
& &+\frac{1}{4}A^m_{\ \;n\La}\;\si_{m\ \;\;c}^{\ \;na}\;A^r_{\ s\Pi}\;\si_{r\ \;\;b}^{\ sc}-\frac{1}{4}(-1)^{\La\Pi}\;A^m_{\ \;n\Pi}\;\si_{m\ \;\;c}^{\ \;na}\;A^r_{\ s\La}\;\si_{r\ \;\;b}^{\ sc}, \label{Curd} \\[6pt]
R^{\dot{a}}_{\ \dot{b}\La\Pi}&\!\!\!\!=\!\!\!\!&\frac{1}{2}(-1)^{\La\Pi}\;A^m_{\ \;n\Pi,\La}\;\si_{m\ \;\;\dot{b}}^{\ \;n\dot{a}}-\frac{1}{2}(-1)^{\La\Pi}\;A^m_{\ \;n\La,\Pi}\;\si_{m\ \;\;\dot{b}}^{\ \;n\dot{a}} \nonumber \\[5pt]
& &+\frac{1}{4}A^m_{\ \;n\La}\;\si_{m\ \;\;\dot{c}}^{\ \;n\dot{a}}\;A^r_{\ s\Pi}\;\si_{r\ \;\;\dot{b}}^{\ s\dot{c}}-\frac{1}{4}(-1)^{\La\Pi}\;A^m_{\ \;n\Pi}\;\si_{m\ \;\;\dot{c}}^{\ \;n\dot{a}}\;A^r_{\ s\La}\;\si_{r\ \;\;\dot{b}}^{\ s\dot{c}}, \label{Cure} \\[6pt]
R^a_{\ m\La\Pi}&\!\!\!\!=\!\!\!\!&R^{\dot{a}}_{\ m\La\Pi}=0. \label{Curf}
\end{eqnarray}

We introduce a strange covariant derivative operator ${\stackrel{\leftarrow}{\mathcal{D}}}_\La$ on superspace that acts only on orthosymplectic-frame indices $\scriptstyle m$, $\scriptstyle a$, $\scriptstyle\dot{a}$ with a {\it Lorentz} connection $A^m_{\ \;n\La}$. Then ${\stackrel{\leftarrow}{\mathcal{D}}}_\mu$ is the space-time covariant derivative operator $D_\mu$ (acting on the right). We have
\begin{equation} \label{siD=0}
\si^m_{\ \ a\dot{a}}\,{\stackrel{\leftarrow}{\mathcal{D}}}_\La=0
\end{equation}
since
\[
\begin{split}
\si^m_{\ \ a\dot{a}}\,{\stackrel{\leftarrow}{\mathcal{D}}}_\La=&\,\si^n_{\ \,a\dot{a}}\,A^m_{\ \;n\La}-\frac{1}{2}\si^m_{\ \ b\dot{a}}\,A^n_{\ r\La}\,\si_{n\ \ a}^{\ \,rb}-\frac{1}{2}\si^m_{\ \ a\dot{b}}\,A^n_{\ r\La}\,\si_{n\ \ \dot{a}}^{\ \,r\dot{b}}  \\[5pt]
=&\,\si^n_{\ \,a\dot{a}}\,A^m_{\ \;n\La}-\frac{1}{2}(\si^m_{\ \ b\dot{a}}\,\si_{n\ \ a}^{\ \,rb}+\si^m_{\ \ a\dot{b}}\,\si_{n\ \ \dot{a}}^{\ \,r\dot{b}})A^n_{\ r\La}  \\[5pt]
=&\,\si^n_{\ \,a\dot{a}}\,A^m_{\ \;n\La}-\eta^{m[n}\,\si^{r]}_{\ \;a\dot{a}}\,A_{nr\La}=0.
\end{split}
\]
By making use of (\ref{siD=0}) we find a simple relation between (\ref{Curb})--(\ref{Curc}) and (\ref{Torb})--(\ref{Torc}) so that the new Riemann supertensor (\ref{Cura})--(\ref{Curf}) can be expressed as
\begin{gather}
R^m_{\ \;n\La\Pi}=(-1)^{\La\Pi}\,A^m_{\ \;n\Pi,\La}-A^m_{\ \;n\La,\Pi}+A^m_{\ \;r\La}A^r_{\ n\Pi}-(-1)^{\La\Pi}\,A^m_{\ \;r\Pi}A^r_{\ n\La}, \label{Cur2a} \\
R^m_{\ \;a\La\Pi}=\si^m_{\ \ a\dot{a}}\,R^{\dot{a}}_{\ \La\Pi}, \quad R^m_{\ \;\dot{a}\La\Pi}=\si^m_{\ \ a\dot{a}}\,R^a_{\ \La\Pi},
\quad R^{\dot{a}}_{\ m\La\Pi}=R^{\dot{a}}_{\ m\La\Pi}=0, \label{Cur2b} \\[5pt]
R^a_{\ b\La\Pi}=\frac{1}{2}R^m_{\ \;n\La\Pi}\,\si_{m\ \;\;b}^{\ \;na},  \quad R^{\dot{a}}_{\ \dot{b}\La\Pi}=\frac{1}{2}R^m_{\ \;n\La\Pi}\,\si_{m\ \;\;\dot{b}}^{\ \;n\dot{a}}.  \label{Cur2c}
\end{gather}
Comparing this with the $k\to\infty$ limit of (\ref{Con1}) we see that the various parts of the new gauge curvature are entirely analogous to the corresponding parts of the new gauge potential. This means that, just like the new gauge potential, the new gauge curvature is super-Poincar\'{e}-algebra valued.

Using (\ref{physE}) and (\ref{physA=0}) we obtain from the torsion (\ref{Tora})--(\ref{Torc}) and Riemann supertensor (\ref{Cur2a})--(\ref{Cur2c})
\begin{gather}
\left.R^m_{\ \;\mu\nu}\right|=D_\mu\,e^m_{\ \;\nu}-D_\nu\,e^m_{\ \;\mu}+\frac{1}{2}\si^m_{\ \ a\dot{a}}(\phi^{\dot{a}}_{\ \mu}\,\phi^a_{\ \nu}+\psi^a_{\ \mu}\psi^{\dot{a}}_{\ \nu})  \label{physCur1} \\[5pt]
\left.R^a_{\ \mu\nu}\right|=D_{[\mu}\phi^a_{\ \nu]}, \qquad \left.R^{\dot{a}}_{\ \mu\nu}\right|=D_{[\mu}\phi^{\dot{a}}_{\ \nu]}, \label{physCur2} \\
\left.R^m_{\ \;n\mu\nu}\right|=\mathcal{R}^m_{\ \;n\mu\nu}, \quad \left.R^m_{\ \;a\mu\nu}\right|=\si^m_{\ \ a\dot{a}}\left.R^{\dot{a}}_{\ \mu\nu}\right|, \quad \left.R^m_{\ \;\dot{a}\mu\nu}\right|=\si^m_{\ \ a\dot{a}}\left.R^a_{\ \mu\nu}\right|, \label{physCur3} \\[5pt]
\left.R^a_{\ b\mu\nu}\right|=\frac{1}{2}\mathcal{R}^m_{\ \;n\mu\nu}\,\si_{m\ \;\;b}^{\ \;na},  \quad R^{\dot{a}}_{\ \dot{b}\mu\nu}=\frac{1}{2}\mathcal{R}^m_{\ \;n\mu\nu}\,\si_{m\ \;\;\dot{b}}^{\ \;n\dot{a}},  \label{physCur4}
\end{gather}
where $\mathcal{R}^m_{\ \,n\mu\nu}$ is the space-time Riemann tensor. The Bianchi identities (\ref{Bian1}) and (\ref{Bian2}) are still satisfied by (\ref{physCur1})--(\ref{physCur4}) and (\ref{physE}).\footnote{In the Wess--Zumino formulation the constraints on the superspace torsion mean the Bianchi identities are no longer identities; rather, they give the field equations of supergravity~\cite{wes1}, and this is the {\it a posteriori} motivation for the constraints.} 

It remains to specify the dynamics of the theory defined by the gauge transformations (\ref{spsg1})--(\ref{spsg4}) and curvatures (\ref{physCur1})--(\ref{physCur4}); we shall do this by writing suitable equations of motion. The requirement for the equations of motion is not that they be covariant under the super-Poincar\'{e} gauge transformations (\ref{spsg1})--(\ref{spsg4}), but that they be gauge covariant when $\Ga^m_{\ \;n\mu}$ is on-shell. Although we have obtained this theory from IOSp$(3,1|4)$ gauge theory, it is fruitless to try to obtain suitable field equations by the procedure of imposing our constraints (\ref{Con1}) and limit $k\to\infty$ on the field equations of IOSp$(3,1|4)$ gauge theory: the latter can be obtained from a ($\Ga^A_{\ \;B\mu}$ on-shell) gauge-invariant action by varying {\it independently} the fields $^A\!E_\La$ and $A^A_{\ B\La}$, but this is inconsistent with the constraints (though not with the limit $k\to\infty$) so that this procedure would not result in super-Poincar\'{e}-covariant equations. 

The appropriate equations of motion are
\begin{gather}
\left.R^m_{\ \;\mu\nu}\right|=0, \label{eqn1} \\
\left.R^m_{\ \;\dot{a}\mu\nu}\right|\, e^\mu_{\ m}=0, \label{eqn2} \\
\mathcal{R}_{n\nu}\,e^n_{\ \mu}\,e^{\nu m}-\left.R^m_{\ \;\dot{a}\mu\nu}\right|\,\phi^{\dot{a}\nu}-\left.R^m_{\ \;a\mu\nu}\right|\,\phi^{a\nu}=0, \label{eqn3}
\end{gather}
where $\mathcal{R}_{n\nu}=\left.R^m_{\ \;n\mu\nu}\right|\, e^\mu_{\ m}$ is the space-time Ricci tensor. Using (\ref{physCur1})--(\ref{physCur3}) in these equations, we see that (\ref{eqn1}) determines the space-time torsion $\mathcal{T}^m_{\ \ \;\mu\nu}$,
\begin{equation} \label{Gaeofm}
\mathcal{T}^m_{\ \ \;\mu\nu}=\sfrac{1}{2}\si^m_{\ \ a\dot{a}}(\phi^a_{\ \mu}\,\phi^{\dot{a}}_{\ \nu}+\phi^{\dot{a}}_{\ \mu}\phi^a_{\ \nu}),
\end{equation}
(\ref{eqn2}) is the gravitino field equation (recall (\ref{phidef2}))
\begin{equation} \label{phyeofm}
\si^m_{\ \ a\dot{a}}\,D_{[\mu}\phi^a_{\ \nu]}\,e^\mu_{\ m}=0,
\end{equation}
and (\ref{eqn3}) is the tetrad field equation
\begin{equation} \label{eeofm}
\mathcal{R}_{\mu}^{\ m}+\si^m_{\ \ a\dot{a}}(\phi^{\dot{a}\nu}\,D_{[\mu}\phi^a_{\ \nu]}+\phi^{a\nu}\,D_{[\mu}\phi^{\dot{a}}_{\ \nu]})=0.
\end{equation}
Transvecting (\ref{eeofm}) with $e^\mu_{\ m}$ and using (\ref{phyeofm}) we obtain
\[
\mathcal{R}^m_{\ \;m}=0,
\]
so that we may replace $\mathcal{R}_{\mu}^{\ m}$ in (\ref{eeofm}) by the space-time Einstein tensor $\mathcal{G}_{\mu m}$. Employing also the the following equivalent form of the gravitino field equation
\[
D_{[\mu}\phi_{\ \nu]}^{a}-\frac{1}{2}i\ep_{\mu\nu}^{\ \ \;\la\rho}\,D_{[\la}\phi_{\ \rho]}^{a}=0 
\]
we can rewrite (\ref{eeofm}) as
\begin{equation} \label{eeofm2}
\mathcal{G}^\mu_{\ \;m}+\frac{1}{2}i\ep^{\mu\nu\la\rho}\,\si_{ma\dot{a}}(\phi_{\ \nu}^{\dot{a}}\,D_\la\phi_{\ \rho}^{a}-\phi_{\ \nu}^{a}\,D_\la\phi_{\ \rho}^{\dot{a}})=0,
\end{equation}
which is the tetrad field equation as it emerges from the supergravity action. As discussed in~\cite{cas}, for example, the action giving rise to (\ref{Gaeofm}), (\ref{phyeofm}) and (\ref{eeofm2}) is invariant under (\ref{spsg1})--(\ref{spsg4}) when (\ref{Gaeofm}) is satisfied. Therefore (\ref{eqn2}) and (\ref{eqn3}) are covariant under (\ref{spsg1})--(\ref{spsg4}) when (\ref{eqn1}) is satisfied.

The geometrical view of simple $D=4$ supergravity presented here is that, although it is not a simple theory of superspace geometry, it is related to a simple theory of superspace geometry (gauge supersymmetry) in a physically understandable way. A comparable theory is Newtonian gravity, which also has a rather complicated geometrical formulation, in terms of space-time~\cite{mis}. Moreover, Newtonian gravity in its space-time form can be derived from general relativity, formulated as Poincar\'{e} gauge theory, by a method analogous to that used here to derive supergravity from IOSp$(3,1|4)$ gauge theory~\cite{phi}. In place of (\ref{Coord}) and (\ref{Met}) one has Minkowski coordinates with $x^0=t$, not $x^0=ct$, so that $\eta_{00}=-c^2$. Instead of (\ref{iosp}) and (\ref{spmat}) one has infinitesimal Poincar\'{e} and Galilean transformations; one obtains the latter from the former by taking the analogous limit $c\to\infty$, but here no constraints are required. One then imposes a similar limiting procedure on the Lorentz gauge potential of Poincar\'{e} gauge theory. In curved space-time one takes a local frame with $\eta_{\widehat{0}\widehat{0}}=-c^2$ and coordinates with $[x^\mu]=-1$ (cf.\ (\ref{Horth}) and (\ref{Zcoord})); a 3-space metric, but not a space-time metric, is preserved when $c\to\infty$~\cite{mis}. The crucial difference from our derivation of supergravity is the lack of constraints, which allows one to obtain a suitable Newtonian field equation by taking the $c\to\infty$ limit of the Einstein field equation.

\section{Conclusions}
We have obtained a new formulation of simple $D=4$ supergravity in terms of the geometry of superspace. This formulation makes clear a relationship between supergravity and the simplest theory of superspace geometry, namely gauge supersymmetry. When one views gauge supersymmetry as IOSp$(3,1|4)$ gauge theory, then supergravity emerges from manipulating quantities in the theory so that IOSp$(3,1|4)$ becomes the super Poincar\'{e} group. This involves imposing constraints and taking a limit.

In thinking about this relationship between supergravity and gauge supersymmetry, it is interesting to note that a similar relationship exists between Newtonian gravity and general relativity. When one views general relativity as Poincar\'{e} gauge theory, and when one takes $\eta_{\hat{0}\hat{0}}=-c^2$, then Newtonian gravity in its space-time form emerges from manipulating quantities in the theory so that the Poincar\'{e} group becomes the Galilean group. This simply involves taking the limit $c\to\infty$; no constraints are required. 

The only difference between the two relationships
\[
\begin{split}
\text{gauge supersymmetry}\ &\rightarrow\ \text{supergravity} \\
\text{general relativity}\ &\rightarrow\ \text{Newtonian gravity} 
\end{split}
\]
as described here, is the occurence of constraints in the former. This difference is of crucial significance for the dynamics, however; the dynamics of supergravity has no simple relation to the dynamics of gauge supersymmetry~\cite{nat1}. Nevertheless, one may say that in the geometrical formulation given here, simple $D=4$ supergravity is more akin to Newtonian gravity than to general relativity. \vspace{5mm}

\noindent
{\bf Acknowlegements} \\
I wish to thank D.\ Hurley and M.\ Vandyck for helpful discussions. This research was supported by Enterprise Ireland grant SC/2001/041.

\end{document}